\documentclass[%
 reprint,
 amsmath,amssymb,
 aps,
]{revtex4-1}

\usepackage{graphicx}% Include figure files
\usepackage{dcolumn}% Align table columns on decimal point
\usepackage{bm}% bold math
\begin{document}

\preprint{APS/123-QED}

\title{Global topology of contact force networks: \\ new insight into shear thickening suspensions}% Force line breaks with \\

\author{Lance E. Edens}
 \altaffiliation{Department of Chemistry, Washington State University}
\author{Sidhant Pednekar}
  \altaffiliation{Benjamin Levich Institute and Department of Chemical Engineering, The City College of New York, Pacific Northwest National Laboratory}
\author{Jeffrey F. Morris}
 \altaffiliation{Benjamin Levich Institute and Department of Chemical Engineering, The City College of New York}
\author{Gregory K. Schenter}
 \altaffiliation{Pacific Northwest National Laboratory}
\author{Aurora E. Clark}
 \altaffiliation{Department of Chemistry, Washington State University}\email{auclark@wsu.edu}
\author{Jaehun Chun}
  \altaffiliation{Pacific Northwest National Laboratory}
  \email{Jaehun.Chun@pnnl.gov}
 
 \date{\today}% It is always \today, today,

\begin{abstract}
Highly concentrated or ``dense" particle suspensions can undergo a sharp increase in viscosity, or shear thickening, under applies stress.
Understanding the fundamental features leading to such rheological changes in dense suspensions is crucial to optimize flow conditions or to design flow modifiers for slurry processing. While local changes to the particle environment under an applied shear can be related to changes in viscosity, there is a broader need to connect the shear thickening transition to the fundamental organization of particle-interaction forces which lead to long-range organization. In  particular, at a high volume fraction of particles, recent evidence indicates frictional forces between contacting particles is of importance. Herein, the network of frictional contact forces is analyzed within simulated two-dimensional shear thickening suspensions. Two topological metrics are studied to characterize the response of the contact force network (CFN) under varying applied shear stress. The metrics, geodesic index and the void parameter, reflect complementary aspects of the CFN: one is the connectedness of the contact network and the second is the distribution of spatial areas devoid of particle-particle contacts. Considered in relation to the variation of the viscosity, the topological metrics show that the network grows homogeneously at large scales but with many local regions devoid of contacts, indicating clearly the role of stress chain growth in causing the large change in the rheological response at the shear thickening transition.

\end{abstract}

\maketitle

%\tableofcontents
\section{\label{sec:level1}Introduction}

\textquotedblleft Dense\textquotedblright suspensions consist of a high volume fraction of particles immersed in a liquid, and can be found in various man-made and natural materials such as muds, ceramics, and chemically extreme environments like those associated with industrial waste storage and processing sites. Controlling the flow of dense suspensions is one of the most critical aspects for processing but it is challenging because these fluids exhibit a wide range of rheological behavior. These rheological properties derive from an interplay of forces between the particles, coupled with the physicochemical characteristics of both the particles and the suspending medium, under far-from-equilibrium shear flow conditions. \par

Suspended particles may generally be subject to hydrodynamic, van der Waals, electrostatic, Brownian, and frictional forces, each of which may predominate depending upon the conditions. In dense Brownian suspensions, a delicate balance between these forces has been shown to be responsible for yielding, shear-thinning, and shear-thickening, depending on shear rate and the volume fraction of particles \cite{Pednekar:2017}. The present work focuses upon non-Brownian suspensions, where the dominant particle forces are expected to be hydrodynamic and frictional \cite{Pednekar:2018, Mari:2014}. The former can be decomposed into a {``}short-range'' hydrodynamic lubrication force that arises from close proximity between particles and the large pressures associated with relative motion of the particle surfaces, and a {``}long-range'' hydrodynamic force. Particle volume fractions approaching the jamming condition have been shown experimentally to lead to increasing importance of frictional forces between two particles that involves direct particle contact and is physically related to roughness of particle surfaces \cite{Lootens:2005,Lin:2016};  previous study by Mari et al. \cite{Mari:2014} demonstrated that inclusion of stress-induced frictional interactions above an onset stress set by repulsive forces between particles allows the rationalization of strong shear thickening in dense non-Brownian  suspensions. The contribution to the viscosity of short-range hydrodynamics relative to friction is larger at smaller volume fraction of particles - specifically, hydrodynamics is found to be dominant up to an approximate volume fraction $\phi = 0.45$ as seen by comparison of experiments with Brownian hard spheres \cite{DHaene:1993} and Stokesian Dynamics simulations \cite{Foss:2000} at the same conditions.  However, at higher solid fraction in the same set of experiments \cite{DHaene:1993}, the behavior was not reproducible without inclusion of further physics. The rate dependence observed is consistent with hydrodynamics overwhelming the stabilizing effects of Brownian motion and repulsive surface forces, driving particles into contact as shear rate increases \cite{Mari:2015,Pednekar:2017}.

\par
 The viscosity of non-Brownian dense suspensions at zero and infinite shear limits has been understood via a volume fraction of particles scaled by a \emph{relevant} maximum packing fraction (e.g., the Krieger-Dougherty formulation \cite{Krieger:1959}). Such a formulation intrinsically captures the frictional nature of the particles through its influence on the maximum packing fraction. A recent study showed that such a scheme can be extended to bidisperse and polydisperse non-Brownian suspensions \cite{Pednekar:2018}. Yield stress of dense suspensions has been studied by rheometry and subsequent simple rheological models that include the Bingham plastic or Herschel-Bulkley fluid models, combined with qualitative scaling arguments and semi-quantitative theories \cite{Russel:1989,Scales:1998,Zhou:2001,Chun:2011}; these studies show the yield stress arises from a collective effect involving both cohesive forces between particles and microstructure of aggregated particles.  

\par
Particle-based numerical simulations such as the noted Stokesian Dynamics (SD) \cite{Foss:2000} and DPD (Dissipative Particle Dynamics) \cite{Pan:2009} have been employed to study the hydrodynamic role in rheology and microstructure of dispersions, and more recently simulation models which include particle contact have been considered to examine the onset of yield stress,  obscuring of shear thickening, and continuous/discontinuous shear thickening \cite{Pednekar:2017,Mari:2014,Boromand:2018}. Importantly, these studies clearly demonstrate a significant role of the network of particles that come into direct contact with each other and interact via frictional force -- the {``}contact force network'' (CFN). These studies have shown that CFN formation is a critical component of the rheological response for both yielding and shear thickening. Furthermore, the shear thickening transition is shown to be coincident with the appearance of a growing number of frictional contacts as the shear rate (or the shear stress) increases \cite{Mari:2014}. More specifically, it was found that, once the contact network forms and spans the suspension, constraints on the motion of particles due to friction generate correlated regions whose large-scale fluctuational motions are much more dissipative than isolated lubricated sphere motions, and in turn result in the high viscosity of the shear thickened state. \par

To date, analysis of the properties of the contact network and their correlations to rheological response has been limited, although the fraction of frictional contacts (i.e., the number of close interaction contacts that are in the frictional state divided by the total number of such interactions) exhibits a direct relation to the shear stress \cite{Mari:2014,Wyart:2014}. Melrose \& Ball \cite{Melrose:2004} focused on hydrodynamic clustering of particles with a polymer coating that resulted in a repulsion as well as a modified lubrication force. While the contact networks were not the focus of their study, they reported that coordination numbers (contacts per particles) were low at low shear rates before suddenly and sharply increasing at the higher rates associated with shear  thickening.

\par
An emerging field of study is the application of network (or graph) theory to explore the topological organization of physical systems. Papadopoulos et al. \cite{Papadopoulos:2018} offers a valuable review of this approach and how it has been applied to granular packing. This review describes how network theory has been developed and employed across multiple disciplines, resulting in a rich set of tools and analysis techniques. Using the tools of  network theory, complex multi-scale (in both length and time) behavior can be related to the patterns of interactions between participants. We use this approach in the present work to study contact force networks, where spherical particles are the nodes and their particle-to-particle frictional contacts are the edges in the network. We are particularly interested in the transmission of forces across system-wide distances and the role this plays in the bulk rheological behavior, with a focus on the viscosity. Such analysis has been performed for granular systems, where network approaches have followed the evolution of the contact network topology under various conditions \cite{Papadopoulos:2018,Kramar:2014,Tordesillas:2010}, such as jamming \cite{Arevalo:2010}. However, contact networks in dense suspensions, under the influence of hydrodynamic interactions, have not been well explored using network theory, and suspensions have the benefit of providing rate-dependent nonequilibrium stready states which are ideal for statistical analysis of network properties and their dependence on the controlled  parameters.
\par
The basic question we consider is what correlating relationships exist between the global network topology and rheological properties.  We then seek to understand how those relationships elucidate the physical behavior. Toward this end, this work introduces a  topological metric that measures the interconnectedness of CFNs as a function of shear stress or shear rate and area fraction of dense suspensions.  We also introduce a topological parameter that measures areas of the 2D network that are void of frictional contacts.  Using these two topological metrics in tandem reveals that the dominant force networks display homogeneous growth across the entire system, but are locally heterogeneous with regions devoid of contacts. We also show that interactions continue to form well past the shear thickening transition, as the CFN becomes more dense and robust.\par

\section{\label{sec:level1}Computational methods}

\subsection{\label{sec:level2}Simulation protocol}

The shear thickening behavior of suspensions was simulated using the method LF-DEM, which employs a 'lubrication flow' (LF) description of hydrodynamic interactions coupled with discrete element modeling (DEM) of particle contacts.  Also included are conservative forces, e.g. electrostatic repulsion \cite{Mari:2015}. This approach successfully reproduces important aspects of both continuous and discontinuous shear thickening. A brief overview of the LF-DEM simulation method is presented here, and  the interested reader can find a detailed description of the tool elsewhere \cite{Seto:2013,Mari:2014,Mari:2015}, including the influence of attractive forces in shear-thickening suspensions \cite{Pednekar:2017}.

\par
LF-DEM is based on the overdamped Langevin equation, with negligible particle inertia, amounting to a force and torque balance for each particle:
\begin{equation}
0={F}_{H}+{F}_{C}+{F}_{R}		\end{equation}
where hydrodynamic forces (\emph{F}\emph{\textsubscript{H}}), contact forces (\emph{F}\emph{\textsubscript{C}}), and repulsive colloidal forces ($ {F}_{R}$) are considered here. A similar torque balance holds.  An electrostatic repulsion is modeled as 
\begin{equation}
{F}_{R}={F}_{0}\mathrm{\exp }\left(-\kappa h\right).
\end{equation}

\noindent The characteristic length scale of the repulsive force is the Debye length which we set to $\kappa^{-1} = 0.02a$, where $a$ is the particle radius. The dominant contribution of hydrodynamic forces in these dense suspensions is captured through the lubrication force (long-range hydrodynamics are neglected) while contact forces are modeled by linear normal and tangential springs that follow the Coulomb friction law:
\begin{equation}
F_{C,tan} \leq \mu F_{C,nor},
\label{eq:Coulomb}
\end{equation}

\noindent where the static friction coefficient $\mu = 1$ in all work presented here. Spring constants are chosen such that the maximum normal and tangential particle deformation is maintained below $0.04a$ and $0.02a$, respectively.\\

In order to avoid shear-induced ordering, a slightly bidisperse suspension having equal area fractions of particles with radii \emph{a} and 1.4\emph{a} was simulated at total area fractions of $\phi_A = 0.70$, 0.72, 0.74, 0.76, 0.78, and 0.79, all in the dense regime.    The simulations were conducted in a quasi-2D environment of a monolayer of 500 total spheres with Lees-Edwards boundary conditions applied to mimic an infinitely extended material. Shear stress-controlled simulations were performed for 30 strain units over a range of shear stress capturing the CST (continuous shear thickening) and DST (discontinuous shear thickening) transitions. Averaged properties of rheological measures (i.e., relative viscosity in our case) were extracted from the simulation output sampled every 0.01 strain after discarding initial transient data ($\sim$2 strain units); the time or strain average is substituted for the ensemble average as this system is found to be ergodic \cite{Mari:2015}.\par 

\subsection{\label{sec:level2}Stress scaling}
%$\eta_0$
Shear stress dependence of the rheology of dense suspensions is only introduced through the presence of an additional force scale besides hydrodynamic forces. In the present work, electrostatic repulsion is the only competing force and thus the repulsive force at contact ($F_0$) is used to rescale the stress for this system. Consequently, shear stress $\sigma$  is reported in dimensionless form through scaling with $\eta_0 \dot{\gamma_0}$, where $\eta_0$ is the viscosity of suspending fluid and $\dot{\gamma_0} = F_0/(6\pi \eta_0 a^2)$.  Suspension viscosity is presented in non-dimensional form as relative viscosity, $\eta_r = \eta/\eta_0$.\par

\subsection{\label{sec:level2}Network analysis}

In application of network (or graph) theory to suspensions, the particles are represented as nodes (or vertices), while edges between nodes denote an interaction. The mathematical representation of any network is the \emph{N $\times$ N} adjacency matrix \textbf{\emph{A}},
\begin{equation}
 {\mathbf{A}}_{ab}=\left\{\begin{array}{c}0\\ 1, \mbox{if } \alpha_{ab }\mbox{ is satisfied} \end{array} \right\}						
\end{equation}

where \emph{N} is the number of participating nodes and $ {\alpha }_{ab}$ is the criterion for establishing an edge between nodes \emph{a} and \emph{b}. In this work, a frictional contact defines an edge and is established when the center-to-center distance between particle pairs (\emph{d}\emph{\textsubscript{ij}}) is equal to the sum of the particles' radii (\emph{a}\emph{\textsubscript{i}} +\emph{a}\emph{\textsubscript{j}}). This contact force network is then further subdivided into two layers: 1) one based on non-sliding (static) friction edges, and 2) one based upon sliding (dynamic) friction edges (Figure 1).  If equation \ref{eq:Coulomb} is satisfied, the contact is described as static and the connection becomes part of the non-sliding network; otherwise the contact is dynamic and becomes part of the sliding network.\par 

\begin{figure}
\begin{center}
\includegraphics[height=2.25in,width=3.4in]{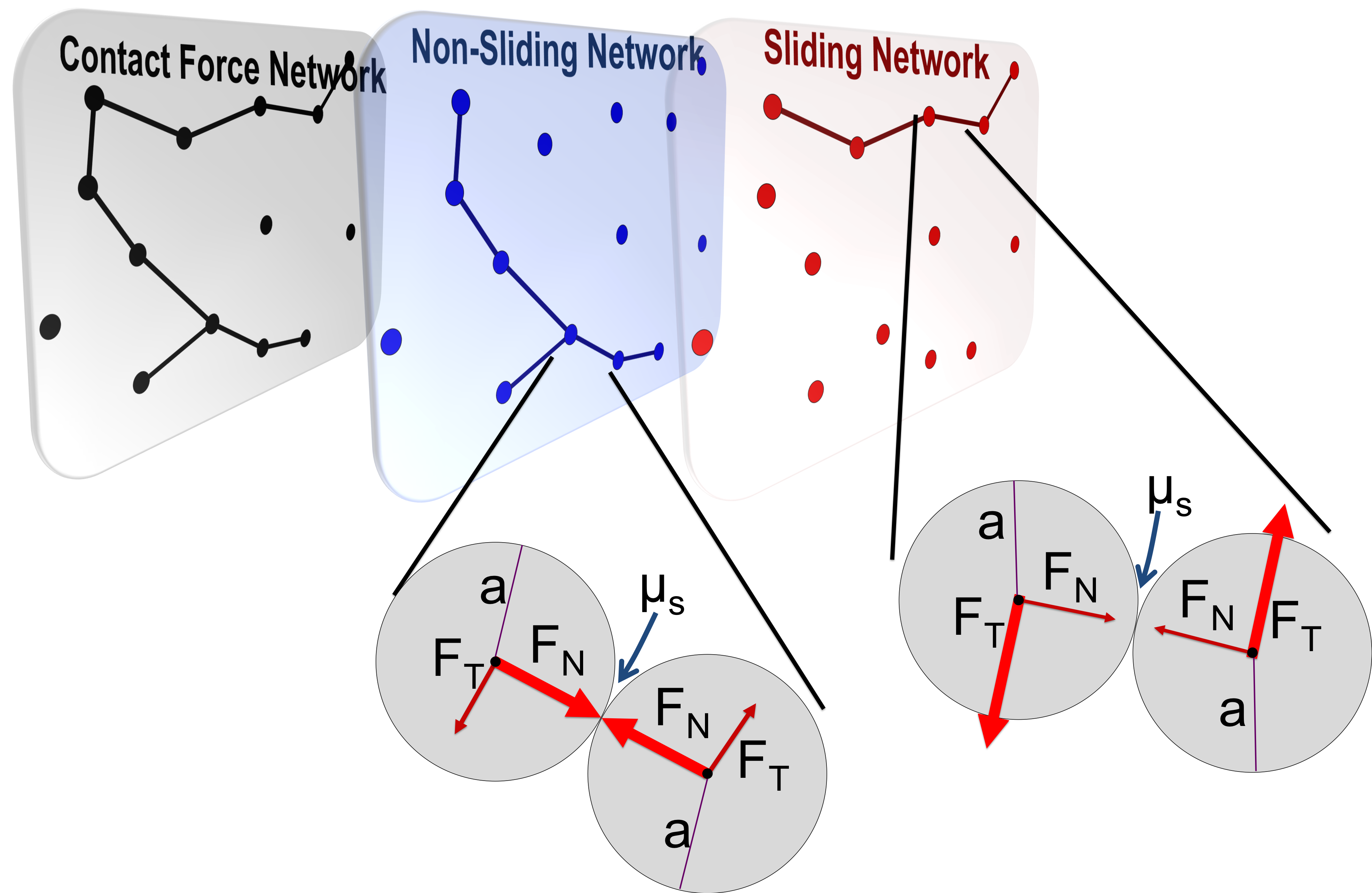}
\caption{A layered contact network, subdivided into the two types of frictional forces: non-sliding and sliding. ${F}_{N}$ and ${F}_{T}$  denote the force normal and tangental to the contact respectively.}
\end{center}
\end{figure}
The simplest analysis of \textbf{\emph{A}} consists of the edge distribution about every node. As prior work has shown \cite{Melrose:2004,Boromand:2018}, the edge distribution of the particles in the CFN (which is a local metric of connectivity) is correlated with the viscosity, presumably because, as viscosity is increased, the maximum local connections possible for a node is reached. For nodes of equal size, the maximum packing arrangement in 2D is a hexagonal lattice, with six contacts per node \cite{Kennedy:2004,Katgert:2010,Saaty:1975}. With two sphere sizes, the ratio of the sphere radii $ {r}_{small}/{r}_{large}$, determines how the nodes will most efficiently pack. In this case, $ {r}_{small}/{r}_{large}> \sqrt{2}-1$ which indicates that, for the bidisperse systems in this work, the smaller spheres cannot fit into the gaps of an ideal hexagonal lattice of larger spheres, and therefore the maximum contact number will be smaller than six \cite{Saaty:1975}.

\begin{figure*}
\begin{center}
\includegraphics[height=1.93in,width=7in]{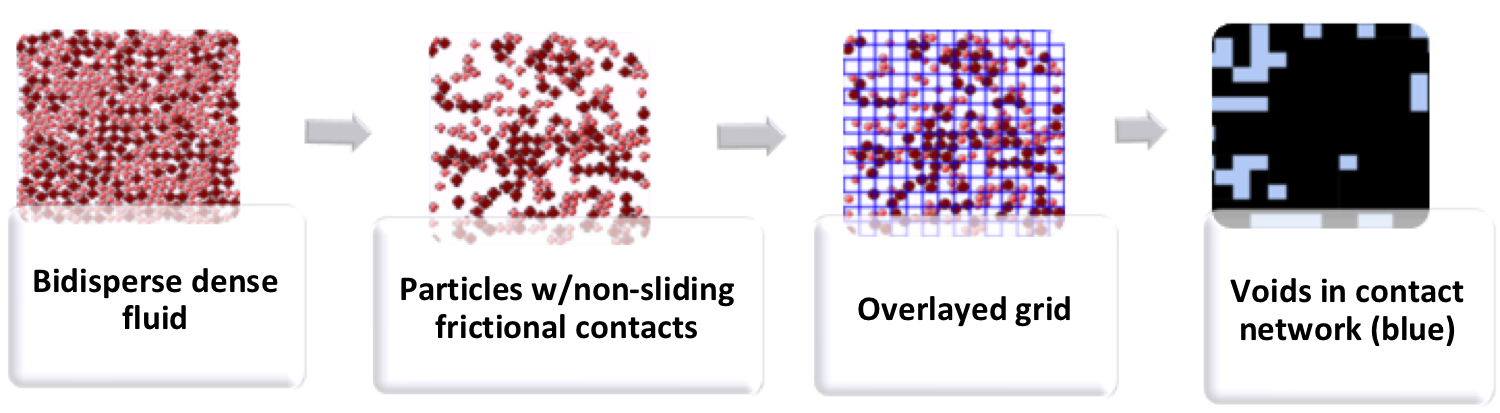}
\caption{Grid analysis to determine void area in non-sliding frictional contact network.}
\end{center}
\end{figure*}

To consider the global topology of the network, we first consider pathways of interactions that span multiple nodes and can have a variety of shapes (cycles, strings, etc.). One topological feature of interest is the extent of connectivity within the network, which can be measured by determining the relative number of paths in which a node participates. It is mathematically expedient to define the shortest pathway that connects every pair of nodes, called the geodesic path. Here, the Floyd-Warshall algorithm \cite{Floyd:1962,Warshall:1962} is used to convert \textbf{\emph{A}} to the geodesic distance matrix containing the shortest number of contiguous edge paths between individual nodes. The raw geodesic distance matrix contains all sub-paths that also connect a pair of nodes. For example, the path from node 1 to node 4 through nodes 2 and 3, the geodesic is 1 $\to$ 2 $\to$ 3  $\to$ 4, yet the entire set of geodesics also contains the sub-geodesics 1 $\to$ 2, 2 $\to$ 3, and 1 $\to$ 2 $\to$ 3.  See Figure S1 in the Supplementary Information. The sub-geodesics are typically removed when considering the number of paths in which a node participates \cite{Clark:2015}. Recognizing that the geodesic matrix is itself size-extensive because the number of geodesics increases with particle count $N$, we  utilize the geodesic index, $\eta_{gd}$, given for a network with $N$ nodes sampled at $M$ snapshots of a simulation by 
\begin{equation}
 {\eta_{gd}= 100 \times [\frac{\sum_{1}^{M}\sigma_{gd(a)}}{N \times M}] }.						
\end{equation}
where
\begin{equation}
 {\sigma_{gd(a)}= \frac{\sum_{1}gd_a(i)+\sum_{1}gd_b(i)+\ldots\sum_{1}gd_N(i)}{N}}						
\end{equation}
and $gd_a(i)$ is the geodesic path in which node $a$ participates. Note that this value can be larger than $N$, as node $a$ can be a linking node in many geodesics that connect other pairs of vertices. Note that the scaling factor of 100 is introduced for convenience. As was shown by Wang et al. \cite{Wang:2014}, a hexagonal ice lattice converges on a geodesic index of about 160. However, this was for a 3D structure of identical-sized nodes. So the value of 160 will not be exceeded by the geodesic index of this system and can safely be considered as a theoretical maximum value. The Floyd-Warshall algorithm, implemented in the ChemNetworks program \cite{Ozkanlar:2014}, was used for all geodesic calculations.

\par
Complementary to the measure of the interconnectedness of the global contact network is the analysis of regions devoid of frictional contacts. To determine the void areas, a grid is applied to each snapshot of the simulation trajectory (illustrated in Figure 2). The grid size is set by the approximate area of two contacting particles, creating a 13 x 13 grid in the relative distance units (scaled to the smallest particle size of 1, with a total box side-length of 52). For a given snapshot, only particles with at least one contact force edge are considered. Each grid element that does not contain a particle is marked as empty. The total void count and area are averaged across snapshots and presented as a function of applied shear stress. Analyzing both the interconnectedness and the empty domains has significant potential to yield new physical insight by elucidating the relative homogeneity  of the network as a function of volume fraction and applied shear stress.

\section{Results and discussion}

\subsection{Benchmarking the simulation data}

The shear thickening behavior of the frictional 2D suspensions is illustrated by plotting the relative viscosity versus shear stress ($\sigma$) for the different $\phi_A$ values in the dense regime (Figure 3). At low shear, repulsive forces dominate over the hydrodynamic forces. The electrostatic repulsion maintains particle surface separation, thereby preventing frictional contacts. The competition between shear and the finite-range repulsive force results in shear thinning behavior at low stresses.\par

\begin{figure}
\begin{center}
\includegraphics[height=2.58in,width=3.4in]{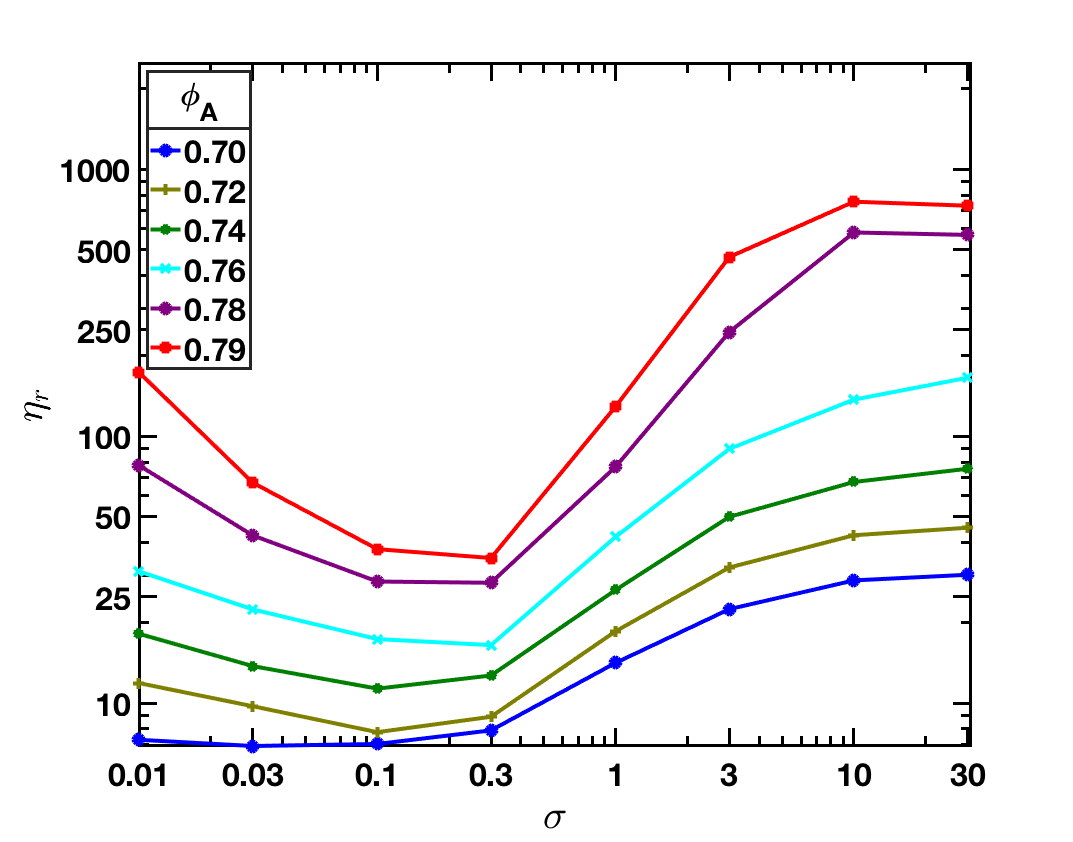}
\caption{Relative viscosities of bidisperse suspensions for the simulations with various $\phi_A$.}
\end{center}
\end{figure}

 Figure 3 shows that onset of shear thickening is observed at a dimensionless critical shear stress of $\sigma\approx0.3$, irrespective of $\phi_A$.  This has been observed in numerous experiments \cite{Frith:1996,Maranzano:2001,Lootens:2005,Larsen:2010,Brown:2012,Brown:2014}. At the onset stress, a non-negligible fraction of the pairwise repulsive forces are overcome by  hydrodynamic shearing to cause particle contact and activate formation of frictional contacts between particles; contact is allowed due to a limiting of the lubrication resistance as described in prior work \cite{Mari:2014}. Frictional forces, and the formation of a frictional contact network, provides additional resistance to flow, leading to increased viscosity observed as either CST or DST depending on $\phi_A$. Phenomenologically, the system experiences a transition from frictionless to frictional rheology at the onset of shear thickening, which is the same mechanism as described for 3D systems \cite{Singh:2018}, and known to compare well with experiments. The 2D suspension is convenient, however, as it allows for straightforward visualization and analysis of network connectivity. Shear thickening in these systems is seen to transition from CST to DST at $\phi_A \approx 0.78$ as indicated by the slope of viscosity as a function of stress approaching the value of 1 \cite{Brown:2010}. This, in 3D simulations, has been known to occur at $\phi\approx0.56$ for friction coefficient of $\mu = 1$ \cite{Mari:2015}.

\subsection{Network connectivity}

The networks analyzed below are a graphical representation of the frictional interactions between particles in non-Brownian dense suspensions. The topological morphologies of these networks are pursued to provide insight to the phenomena responsible for rheological properties. Exploring changes in graphical network parameters allows us to determine the relationship between frictional force and networks across length-scales (local to global topology) and bulk shear-thickening.\par

\begin{figure}
\begin{center}
\includegraphics[height=5.45in,width=3.4in]{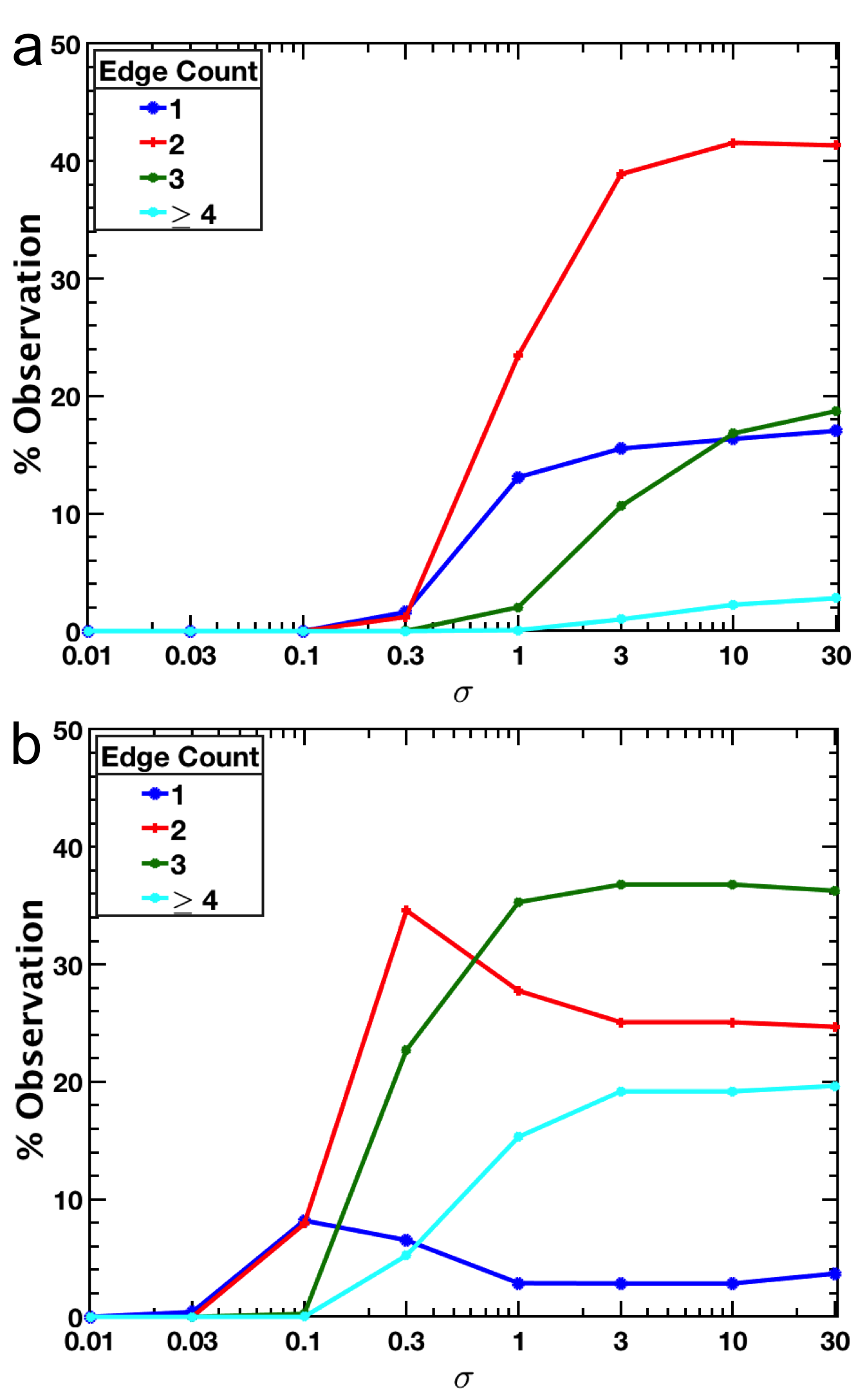}
\caption{Distribution of direct particle contacts for the non-sliding contact networks from (a) $\phi_A$=0.70 and (b) $\phi_A$=0.79.}
\end{center}
\end{figure}

The edge distribution over all nodes in the contact force network is a local measure of connectivity. Figure 4 presents the edge distributions of non-sliding force networks at the lowest and highest area fractions studied, $\phi_A = 0.70$ and 0.79, with intermediate fractions presented in supplementary information. As anticipated, the system with a higher $\phi_A$ has a larger population of particles with more contacts. Further, at the large $\phi_A$, the number of contacts increases rapidly as shear stress increases, even when the shear stress is at the onset value of $\sigma\approx0.3$. For all $\phi_A$ values, edge connections reach a plateau around a shear stress of $\approx 5$, but the viscosities continue to increase with increasing $\sigma$ (see Figure 3). Note that if the average number of contacts is considered, the value for all $\phi_A$ values plateaus to approximately 3 contacts and is thus well below what would be anticipated for a hexagonal (edge count of 6) 2D system. This implies that even at the highest shear stress, approaching a jammed state, the system is far below the maximum number of contacts for an ordered system. While this local connectivity information reveals an average of the local particle interactions, the global rheological characteristics are not captured solely by the local contacts of the particles, providing motivation for study of the topological characteristics of the entire contact network and their changes under applied shear stress.\par

In contrast to the local structure of the network provided by the edge distribution described above, the geodesic index is a global metric of the extent of interconnectivity of the CFN. The geodesic index as a function of shear stress is shown in Figure 5 for $\phi_A$ studied, for non-sliding  (Figure 5a) and sliding frictional contact networks (Figure 5b). For the non-sliding network, the geodesic index increases with shear stress once past the shear thickening onset of $\sigma\approx0.3$. Below this value, the geodesic index remains small due to the low number of frictional contacts; the force networks that the index is measuring have yet to form in  large numbers. Insensitivity of the geodesic index to the change in viscosity below a $\sigma \approx 0.30$ is a result of the small contribution of the non-sliding contact force to the suspension stress. For $\sigma> 0.3$, the geodesic index begins to rise, but at different rates depending on $\phi_A$. As with the viscosity, the increase in the geodesic index is faster for the higher $\phi_A$. The dependence on $\phi_A$ is also seen in the rate at which the geodesic index approaches its saturation value, which occurs when the maximum number of particles for a specific $\phi_A$ are participating in the network. In contrast, the sliding frictional network, while containing fewer connections than its non-sliding counterpart, begins to increase its connectivity at a lower shear stress. At higher stress, the curves turn over and the sliding network connections begin to decrease, with higher $\phi_A$ decreasing faster. This is similar to the edge count distributions, indicating that the sliding frictional contacts are likely confined to localized networks. \par

Our analysis of these networks provides a picture of the type of interactions in which the nodes (i.e. particles) participate. Below the onset stress for shear thickening of $\sigma \approx 0.3$, the stress is primarily generated by hydrodynamic forces and weak shear thickening is seen \cite{Mari:2014,Wagner:2009}. Just prior to the critical stress, the interconnectivity of the nodes begins to grow, with the dynamic frictional contacts undergoing rapid growth as the stress begins to overcome the repulsion force. At the onset stress, networks of non-sliding frictionally-contacting particles rapidly grow, and these apparently are crucial to the more rigid correlated motion leading to elevated viscosity. With further increase in $\sigma$, the non-sliding networks become system-spanning; the detailed conditions for the system spanning network are affected by the finite simulation size, and we have not probed these here.\par
\begin{figure}
\begin{center}
\includegraphics[height=5.66in,width=3.4in]{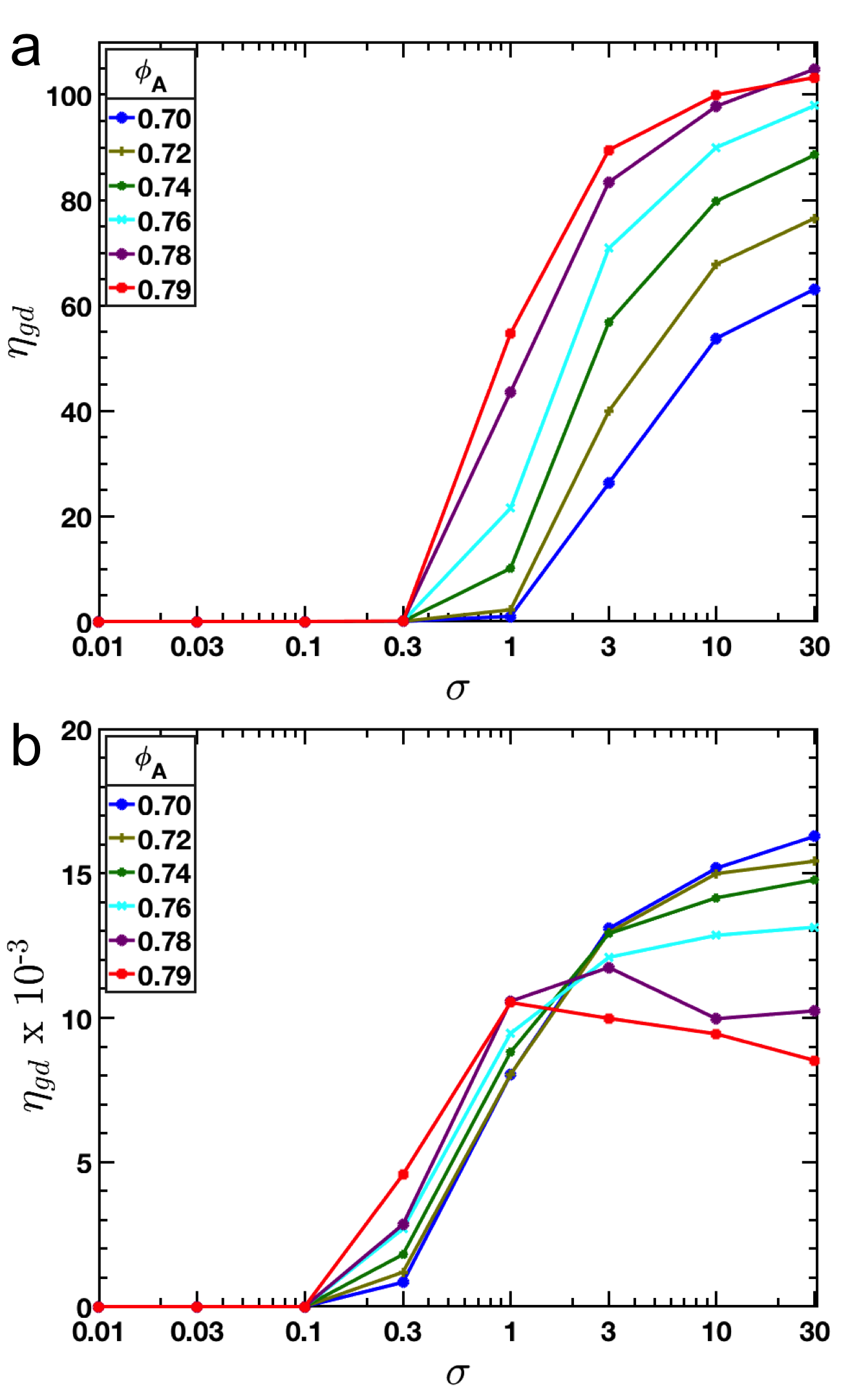}
\caption{Changes in the geodesic index for a) non-sliding contact force networks and b) sliding contact force networks. The non-sliding network with error bars included is shown in the Supplementary Information.}
\end{center}
\end{figure}

\subsection{Voids: Domains that lack particle contacts}

Complementary to understanding the interconnectedness of the contact network is a depiction of the unoccupied regions that remain as stress increases. This metric conveys information on the patterns and shapes associated with the 2D area of the forming networks. Thus, these empty domains or “voids” offer a measure of the developing force network, as the force chains form the boundaries of the voids in the highly connected states. A region of high connection density will produce many small voids. The total void area and the number of separate void domains for the non-sliding force networks are presented for different $\phi_A$ in Figure 6 as a function of shear stress. The sliding network void analysis is included in the Supplementary Information. Though beyond the scope of the analysis presented here, more detailed analysis of the 2D void shapes could provide added insight into the change in rheological properties. Note that the void analysis can be extended to 3D, where the voids would be volumes. 

\begin{figure}
\begin{center}
\includegraphics[height=5.52in,width=3.4in]{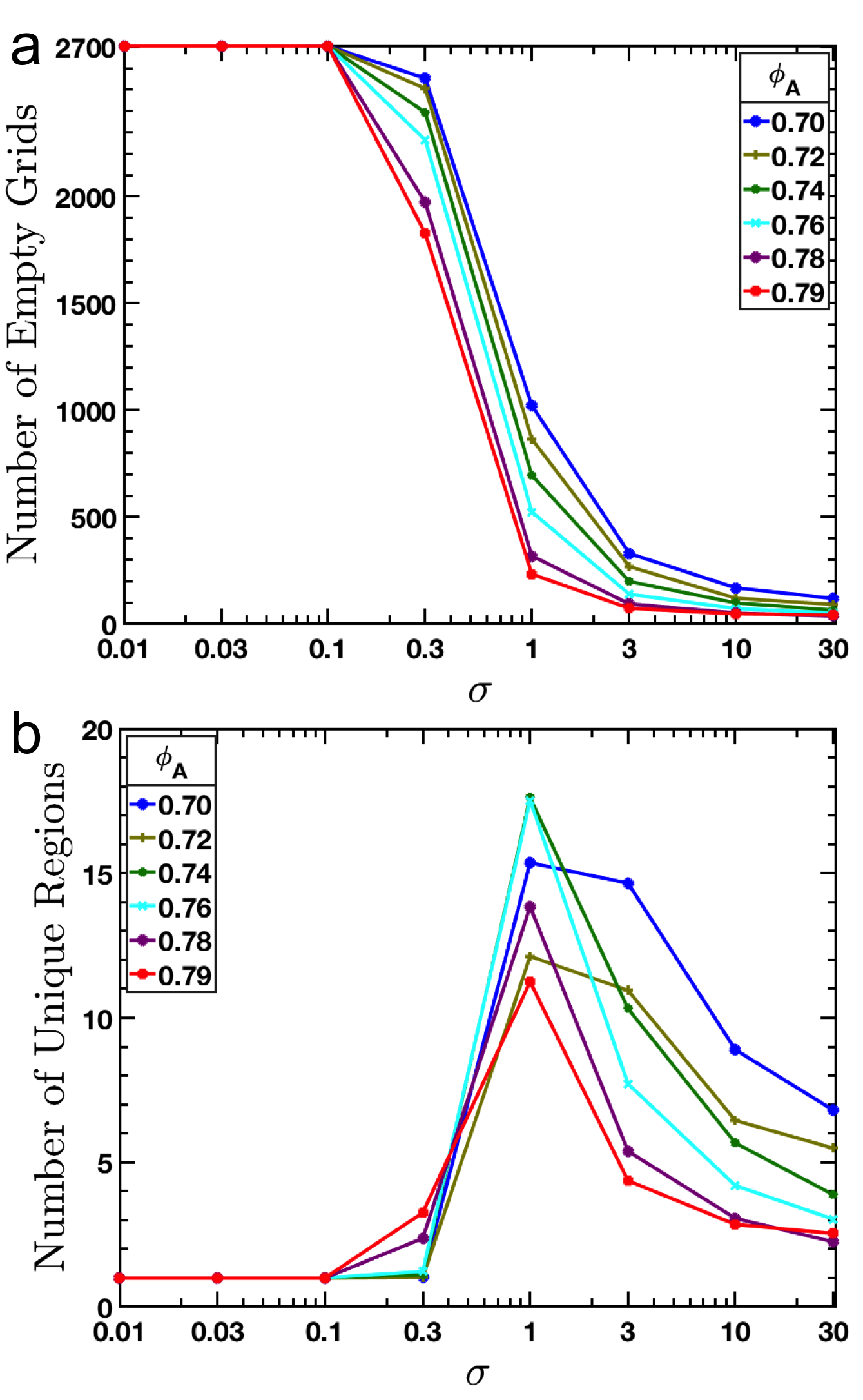}
\caption{Void domain analysis for non-sliding force networks. (a) The number of empty grid squares (void domains). (b) Number of distinct void domains. All curves were averaged over the snapshots of the simulation.}
\end{center}
\end{figure}

\par
Initially there are no contacts, and the entire domain is therefore a single void. When contacts begin to occur, at $\sigma = 0.1 - 0.3 $ depending on $\phi_A$, the number of voids remains small and consequently the void areas.  Both features are insensitive to the changes in viscosity over this initial evolution where hydrodynamic stress remains dominant, similar to the geodesic index. Visually, we observe that most of the simulation plane is empty of contact network connections at low shear stress, and the few void areas present span most of the domain. As the shear stress increases, networks are formed as previously demonstrated by the growth in the number of edges, and the early rise in the geodesic index. Once past the onset stress of $\sigma \approx 0.3$, all volume fractions show a rapid increase in the distinct number of void regions. As the network forms, space is effectlvely partitioned by the force network into separate void domains, so that the average area of the voids falls rapidly with increase of $\sigma$. Examples of the separating void domains are shown in Figure 7. 
\begin{figure}
\begin{center}
\includegraphics[height=2.93in,width=3.4in]{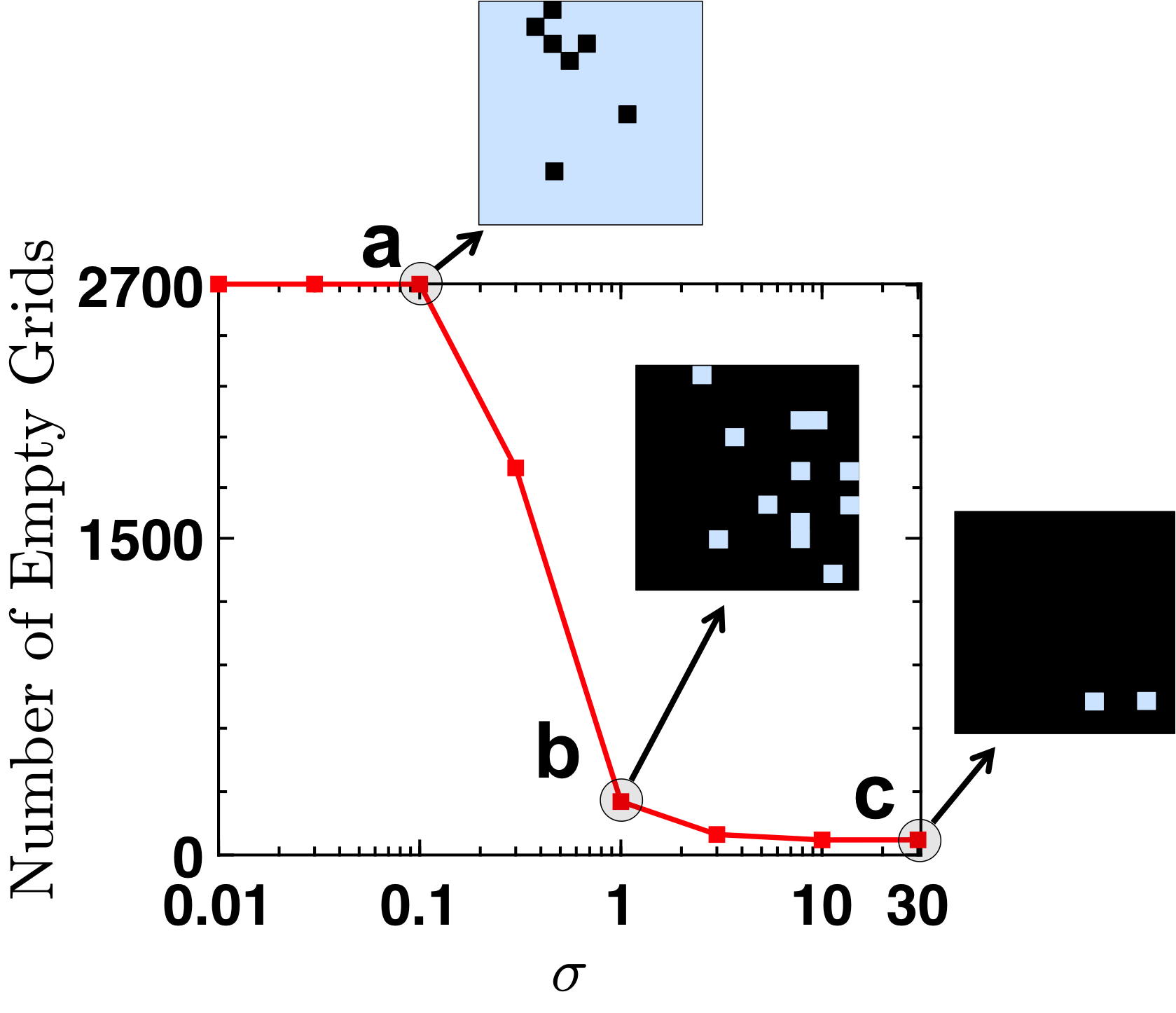}
\caption{Examples of void domains (blue) for ϕ = 0.79. (a) A snapshot of a large single void domain at low shear stress. (b) A snapshot showing several smaller void domains in intermediate shear stress. (c) A snapshot of very small void domains in high shear stress.}
\end{center}
\end{figure}

\par
Interestingly, for stress above $\sigma \approx 1.0$, the void counts of the non-sliding networks in Figure 6b decrease, at a rate depending on the area density. This occurs even while the geodesic index continues increasing quite rapidly, implying there is a limit to the number of small voids that may be created within the network. As the network connections continue to span across void domains, many of the empty regions within the network vanish (i.e. become smaller than $2a$). This effect is density-dependent as the smaller void domains are eliminated at lower stress for higher $\phi_A$. However, even at the maximum $\sigma$, some small voids always remain. This is consistent with the disorder of the structure, as the edge distribution never approaches the hexagonal packing limit. Some regions are not part of the contact network even in the high viscosity state.

\subsection{Complementary frameworks: interconnectedness and voids }

The rapid increase in the void count and the decrease in void area suggest that network formation for non-sliding frictional contacts is relatively homogeneous and well distributed throughout the domain. If the network grew from a single region, the original void domain in Figure 6b would not separate into subdomains quickly, and total void area presented in Figure 6a would slowly decrease as the networks connections gradually absorbed empty domains. Likewise if the force networks formed only on the borders of the empty domains and worked their way into the empty regions, the decrease in void area and increase in void counts would gradually change. However, we find instead that  the large void areas are quickly spanned and subdivided by the force networks. 
%The formation of frictional contacts is not altered by the presence of the %empty domains, and network formation proceeds at the same rate across the %entire simulation.
% [I could not fully understand the first part of the sentence and thought the 
%point at end was well-made alerady -- Jeff ]
\par
When considered together, we see that the geodesic index and the void analysis exhibit different behaviors at the onset of the discontinuous shear thickening transition, during and after the transition.  As the non-sliding frictional network is initially formed, the void count and area change dramatically with the rapid rise in the interconnectivity until the network spans the system. For this regime, the analysis of the voids becomes particularly insightful. However, as the viscosity continues to rise, both the number of voids and their area are effectively minimized, and it is the interconnectedness of the contact network itself that is the most sensitive to the rise in viscosity. 

\section{Conclusions}

A quantitative understanding, of non-Newtonian behavior of fluids requires knowledge of the fundamental forces and their underpinning organization. Here, we introduce and evaluate global topological parameters of the contact force network of a shear thickening fluid. We show that these can not only track changes in rheological properties (demonstrated for viscosity), but also reveal important insights during the transition in properties. The network analyses introduced here for the suspension behavior --  namely geodesic index and void count -- illustrate that discontinuous shear transition is characterized by: 1) a rapid rise in the interconnectivity of the contact force network, and 2) the homogeneous distribution of void domains whose number and area are minimized uniformly throughout the domain by the enhanced interconnectivity. These well-defined network topological parameters provide a new window to understand the coupled behavior of the particle forces and, in particular, how the global features of the network can be correlated to the  bulk rheological responses. The scheme can help articulate a possible correlation between parameters used in empirical viscosity models (e.g., Krieger-Dougherty equation) and the contact network (and thus particle forces).\par

\section{Acknowledgements}
This research was supported by the Interfacial Dynamics in Radioactive Environments and Materials (IDREAM), an Energy Frontier Research Center funded by the U.S. Department of Energy, Office of Science, Basic Energy Sciences. 

\bibliography{2DNetworkPaper_edifix_BibTeX.bib}

\end{document}